\documentclass[fleqn,usenatbib]{mnras}

\usepackage{newtxtext,newtxmath}

\usepackage[T1]{fontenc}
\usepackage{ae,aecompl}

\usepackage{graphicx}	
\usepackage{amsmath}	
\usepackage{amssymb}	
\let\plate\undefined
\usepackage{tikz}
\usetikzlibrary{bayesnet}

\newcommand{\popp}{\boldsymbol{\Psi}}
\newcommand{\objp}{\boldsymbol{\psi}}
\newcommand{\data}{\mathbf{d}}
\newcommand{\Teff}{T}
\newcommand{\logg}{g}
\newcommand{\pr}{\text{Pr}}
\newcommand{\de}{\text{d}}
\newcommand{\kpc}{\text{kpc}}
\newcommand{\K}{\text{K}}

\title[Inferring properties of the white dwarf population]{Inferring properties of the local white dwarf population in astrometric and photometric surveys}

\author[A. Widmark et al.]{
Axel Widmark,$^1$\thanks{E-mail: axel.widmark@fysik.su.se} 
Daniel J. Mortlock,$^{2,3,4}$
Hiranya V. Peiris$^{1,5}$
\\
$^1$The Oskar Klein Centre for Cosmoparticle Physics, Department of
Physics, Stockholm University, AlbaNova, 10691 Stockholm, Sweden\\
$^2$Astrophysics Group, Imperial College London, Blackett Laboratory, Prince Consort Road, London SW7 2AZ, UK\\
$^3$Department of Mathematics, Imperial College London, London SW7 2AZ, UK\\
$^4$Department of Astronomy, Stockholm University, AlbaNova, SE-10691 Stockholm, Sweden\\
$^5$Department of Physics and Astronomy, University College London, Gower Street, London, WC1E 6BT, UK\\
}

\date{Accepted XXX. Received YYY; in original form ZZZ}

\pubyear{2018}

\begin{document}
\label{firstpage}
\pagerange{\pageref{firstpage}--\pageref{lastpage}}
\maketitle

\begin{abstract}
The \emph{Gaia} mission will provide precise astrometry for an unprecedented number of white dwarfs (WDs), encoding information on stellar evolution, Type Ia supernovae progenitor scenarios, and the star formation and dynamical history of the Milky Way. With such a large data set, it is possible to infer properties of the WD population using only astrometric and photometric information. We demonstrate a framework to accomplish this using a mock data set with SDSS \emph{ugriz} photometry and \emph{Gaia} astrometric information.
Our technique utilises a Bayesian hierarchical model for inferring properties of a WD population while also taking into account all observational errors of individual objects, as well as selection and incompleteness effects.
We demonstrate that photometry alone can constrain the WD population's  distributions of temperature, surface gravity and phenomenological type, and that astrometric information significantly improves determination of the WD surface gravity distribution. We also discuss the possibility of identifying unresolved binary WDs using only photometric and astrometric information.
\end{abstract}




\section{Introduction}

White dwarfs (WDs) are the remnants of stars with initial masses $\lesssim8\mbox{--}10$~M$_\odot$ \citep{1996ApJ...460..489R,2009MNRAS.395.1409S}. The local WD population carries information about the Galaxy's star formation and dynamical history, and constrains models of stellar evolution \citep{1987ApJ...315L..77W,2016NewAR..72....1G,2018arXiv180505849E}.

The Sloan Digital Sky Survey (SDSS) catalogued ${\sim}29,000$ spectroscopically confirmed WDs \citep{2013ApJS..204....5K,2015MNRAS.446.4078K}. A fundamental difficulty in studying WDs is that their mass is degenerate with distance. This degeneracy can be broken with high quality spectrometry and accurate atmospheric models. The \emph{Gaia} mission, which recently published its second data release (DR2), is expected to increase the number of known WDs by approximately an order of magnitude \citep{Jordan:2006jg,2014A&A...565A..11C}; \cite{2018MNRAS.tmp.1537K} and \cite{2018arXiv180703315G} have recently published WD catalogues, the latter containing 260,000 high-confidence WDs. \emph{Gaia} also provides astrometric information for local neighborhood WDs. For comparison, the astrometric mission \emph{Hipparcos} had a limiting apparent magnitude of $V \sim 12.4$ \citep{1997A&A...323L..49P}, while \emph{Gaia} will see objects as faint as $G \sim 21$ (with this limit, a WD with a mass of 0.6 M$_\odot$ and effective temperature of 8,000 K is seen to ${\sim}400$ pc, assuming no dust extinction).

In a model of the WD population, it is physically meaningful to divide the total population into WD sub-populations. WDs form a family of atmospheric types, where the main classification is between DA and DB, depending on if the envelope is hydrogen- or helium-dominated \citep{Tremblay:2007hq,2011ApJ...737...28B,2015A&A...583A..86K}. DA and DB stars can be identified with accurate photometry, as demonstrated by \cite{Mortlock:2008gf}. The halo WD population is kinematically warmer and older than the disk WD population, such that inferring and comparing properties of these sub-populations can yield information on the star formation and kinematic history of our Galaxy \citep{1998ApJ...503..239I,2016MNRAS.463.2453D}. The sub-population of binary WD systems holds information about stellar evolution \citep{Postnov:2014tza} and Type Ia supernovae progenitor scenarios \citep{Livio:2018rue}, but unresolved binaries are very difficult to identify even with high-quality spectroscopy.

With the enormous size of the \emph{Gaia} data set, there is great potential for inferring properties of the WD population using photometry and astrometry, rather than the smaller data set of spectroscopically observed WDs. In this work we demonstrate how to infer properties of the WD population in the solar neighborhood, using SDSS \emph{ugriz} photometry and \emph{Gaia} astrometry. We generate a mock data sample of WDs from a population model of temperature, surface gravity, and spatial number density distribution, of DA and DB atmospheric types. All sample objects have photometry and parallax information with observational errors expected from SDSS and \emph{Gaia}, and sample construction selection effects are taken into account. We also discuss the possibility of identifying binary WD systems and demonstrate how to do so using photometric and astrometric information alone.

This paper is organized as follows. We outline our model for the WD population and the observational data that we consider, in Sec.~\ref{sec:model} and Sec.~\ref{sec:data} respectively. We present out method of statistical inference in Sec.~\ref{sec:method}.  followed by Sec.~\ref{sec:mock}, where we generate a mock data catalogue and infer the model parameters from that data. We discuss possible extensions to the WD model in Sec.~\ref{sec:subpopulations}, such as differentiating between disk and halo sub-populations, as well as the possibility of identifying unresolved double-degenerate binary WD systems. Finally, in Sec.~\ref{sec:discussion} we present our conclusions.

\section{Model}\label{sec:model}

We model the Milky Way's population of WDs by a spatial distribution, and distributions in intrinsic WD properties. A WD is parametrized by effective temperature ($\Teff$), surface gravity ($\logg$), phenomenological type ($t$), and spatial position ($\mathbf{x}$). These are listed in Table \ref{tab:parameters}, along with the population parameters and data.

There are five population parameters in our model, encapsulated in a vector $\popp$: the population parameter $\alpha$, which parametrizing the distribution of temperatures; $\bar{g}$, $\sigma_g$, $\gamma_g$ which parametrize the distribution of surface gravities; and $f_\text{DB}$ which is the fraction of DB-type WDs.

The distribution of effective temperature is parametrized as
\begin{equation}\label{eq:T}
    \pr(\Teff | \popp) \propto \Theta(\Teff - \Teff_\text{min})\, \Theta(\Teff_\text{max} - \Teff) \, \exp (-\alpha \Teff),
\end{equation}
where $\Theta$ is the Heaviside step function, and $\Teff_\text{min}=1500$~K and $\Teff_\text{max}=120$,000~K the lower and upper bounds to the effective temperature.

The distribution of surface gravity is parametrized by
\begin{equation}\label{eq:T&g}
    \pr(\logg | \popp) \propto \Theta(\logg - \logg_\text{min}) \, \Theta(\logg_\text{max} - \logg) \, f_t(\logg|\bar{g},\sigma_g,\gamma_g),\\
\end{equation}
where $\logg_\text{min}=7$ and $\logg_\text{max}=9$ are the lower and upper bounds to the surface gravity, and $f_t(\logg)$ is Student's $t$-distribution of width $\sigma_g$ and variance $\text{Var}(g) = \gamma_g^2 \sigma_g^2$, such that the quantity $\gamma_g$ characterizes the heaviness of the distribution's tails.
\footnote{
Student's $t$-distribution is defined
\begin{equation}
	f_t(\logg|\bar{g},\sigma_g,\nu) =
	\frac{\Gamma\big( \frac{\nu+1}{2} \big)}{\sqrt{\pi\sigma_g^2\nu}\;\Gamma\big( \frac{\nu}{2} \big)}
	\Bigg( 1+\frac{(\logg-\bar{g})^2}{\sigma_g^2\nu} \Bigg)^{-\frac{\nu+1}{2}},
\end{equation}
where we reparametrize the shape parameter according to
\begin{equation}
	\gamma^2 = \frac{\nu}{\nu-2}.
\end{equation}
}

The type of the object constitutes a third parameter of the intrinsic WD properties, called $t$. This is a label, denoting for example if the WD is of DA or DB atmospheric classification. The probabilities are written in terms of the fraction of DB stars, as
\begin{equation}\label{eq:DADB}
\begin{split}
	\pr(t=\text{DA} | \popp) & = (1-f_\text{DB}),\\
    \pr(t=\text{DB} | \popp) & = f_\text{DB}.
\end{split}
\end{equation}

Given the intrinsic properties of a WD, the absolute magnitude in different photometric bands can be calculated using a stellar model. In this paper, we use the Bergeron atmospheric models for WDs \citep{Bergeron:1995we,Finley:1997zz,Bergeron:2000ce,2001PASP..113..409F}.

Also included in our model is a WD number density function, based on a Galactic model by \cite{2008ApJ...673..864J}, expressed in terms of cylindrical coordinates $R$, $Z$ and $\Phi$:
\begin{equation}\label{eq:numberdensity}
\begin{split}
	n(\mathbf{x}) \propto
	\Bigg\{ 
		& \exp\Bigg(-\frac{R}{L_\text{thin}}\Bigg)\exp\Bigg(-\frac{|Z|}{H_\text{thin}}\Bigg) \\
		& +f_\text{thick}\exp\Bigg(-\frac{R}{L_\text{thick}}\Bigg)\exp\Bigg(-\frac{|Z|}{H_\text{thick}}\Bigg) \\
		& +f_\text{halo}\Bigg[ \frac{(R^2+Z^2/q_\text{halo}^2+R_\text{core}^2)^{1/2}}{L_\text{halo}} \Bigg]^{-\eta_\text{halo}}
	\Bigg\},
\end{split}
\end{equation}
where $f_\text{thick}=0.06$, $f_\text{halo}=6\times10^{-5}$, $L_\text{thin}=L_\text{thick}=3.5~\kpc$, $L_\text{halo}=8.5~\kpc$, $R_\text{core}=1~\kpc$, $H_\text{thin}=0.26~\kpc$, $H_\text{thick}=1~\kpc$, $q_\text{halo}=0.64$ and $\eta_\text{halo} = 2$. Assuming a solar position of $R_\odot=8~\kpc$ (where the height of the Sun above the plane is neglected), the Galactic coordinates are given by cylindrical heliocentric coordinates through
\begin{equation}
\begin{split}
	& R(d,l,b) = (d^2\cos^2b-2 R_\odot d \cos^2b\cos^2+R_\odot^2)^{1/2}, \\
	& Z(d,l,b) = d \sin b.
\end{split}
\end{equation}
The azimuthal angle can be neglected, as the Galaxy is assumed to be axisymmetric in this model.

\begin{table}
	\centering
	\caption{The population parameters of our model, and the object parameters and data of the respective stars.}
	\label{tab:parameters}
    \begin{tabular}{l l}
		\hline
		$\popp$  & Population parameters \\
		\hline
		$\alpha$ & slope of temperature distribution \\
		$\bar{g}$ & mean surface gravity $\logg$ \\
		$\sigma_g$ & width of $\logg$ distribution \\
		$\gamma_g$ & heaviness of the tails of the $\logg$ distribution \\
		$f_\text{DB}$ & the fraction of DB type WDs \\
        \hline
        $\objp_{i=1,...,N}$  & Object parameters \\
        \hline
        $\Teff$ & effective temperature \\
        $\logg$ & surface gravity \\
        $t$ & phenomenological type (DA or DB) \\
        $\mathbf{x}$ & spatial position  \\
        \hline
        $\data_{i=1,...,N}$ & Data \\
        \hline
        $\hat{m}_b$ & observed photometric magnitude \\
        $\sigma_b$ & magnitude uncertainty \\
        $\hat{\varpi}$ & observed parallax \\
        $\sigma_{\hat{\varpi}}$ & parallax uncertainty \\
        $l$ & observed Galactic longitude \\
        $b$ & observed Galactic latitude \\
		\hline
	\end{tabular}
\end{table}

\section{Data}\label{sec:data}

The data for a given WD are apparent magnitude measurements in photometric bands ($\hat{m}_b$), a parallax measurement ($\hat{\varpi}$), angular position ($l$ and $b$), including the error models of these observables.\footnote{A hatted quantity (e.g. $\hat{\varpi}$) refers to an observed value, while a non-hatted quantity (e.g. $\varpi$) refers to an observable's true value. The angles $l$ and $b$ are written without hats, as their measurement uncertainties are so small that they can be neglected.} The data characterizing a WD with index $i$, called $\data_i$, is listed in Table \ref{tab:parameters}.

The likelihood associated with a WD is
\begin{equation}\label{eq:likelihood}
	\pr(\data_i | \objp_i) = \mathcal{N}(\varpi(\objp_i)|\hat{\varpi},\sigma_{\hat{\varpi}})\prod_{b} \mathcal{N}(m_b(\objp_i)|\hat{m}_b,\sigma_b),
\end{equation}
where $\mathcal{N}(x | \mu,\sigma)$ is the normal distribution of mean $\mu$ and standard deviation $\sigma$. The factor containing parallax information is dropped when no parallax information is present. The apparent magnitudes $m_b(\objp_i)$ are functions of the object parameters, coming from a stellar model.

\begin{figure}
	\includegraphics[width=\columnwidth]{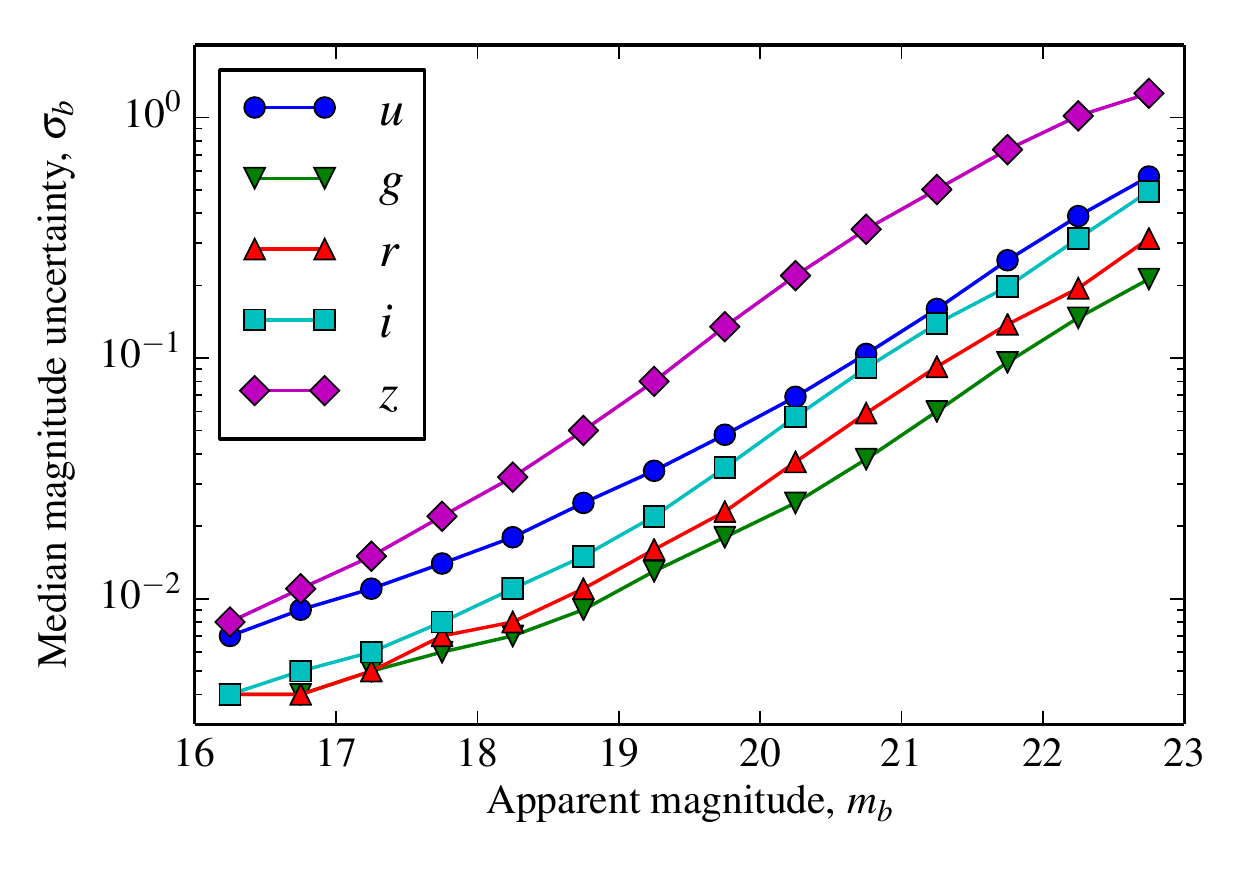}
    \caption{Median magnitude uncertainties for photometric bands $b=\{u,g,r,i,z\}$ in SDSS DR9, as a function of observed apparent magnitude. This is described in greater detail in Sec. \ref{sec:data}.}
    \label{fig:magnitude_error}
\end{figure}

In this work, we use SDSS photometry in \emph{ugriz} colour bands, and a Bergeron atmospheric stellar model, as discussed in Sec. \ref{sec:model}. In order to assign realistic uncertainties to the mock data that we generate, we use median uncertainties of the SDSS DR9 catalogue, in bins of observed apparent magnitude of width 0.5 mag. These median uncertainties are shown in Fig.~\ref{fig:magnitude_error}. We limit the minimum \emph{ugriz} magnitude uncertainty to 0.01 mag, in order to account for possible systematic uncertainties in the Bergeron atmospheric model.

\begin{figure}
	\includegraphics[width=\columnwidth]{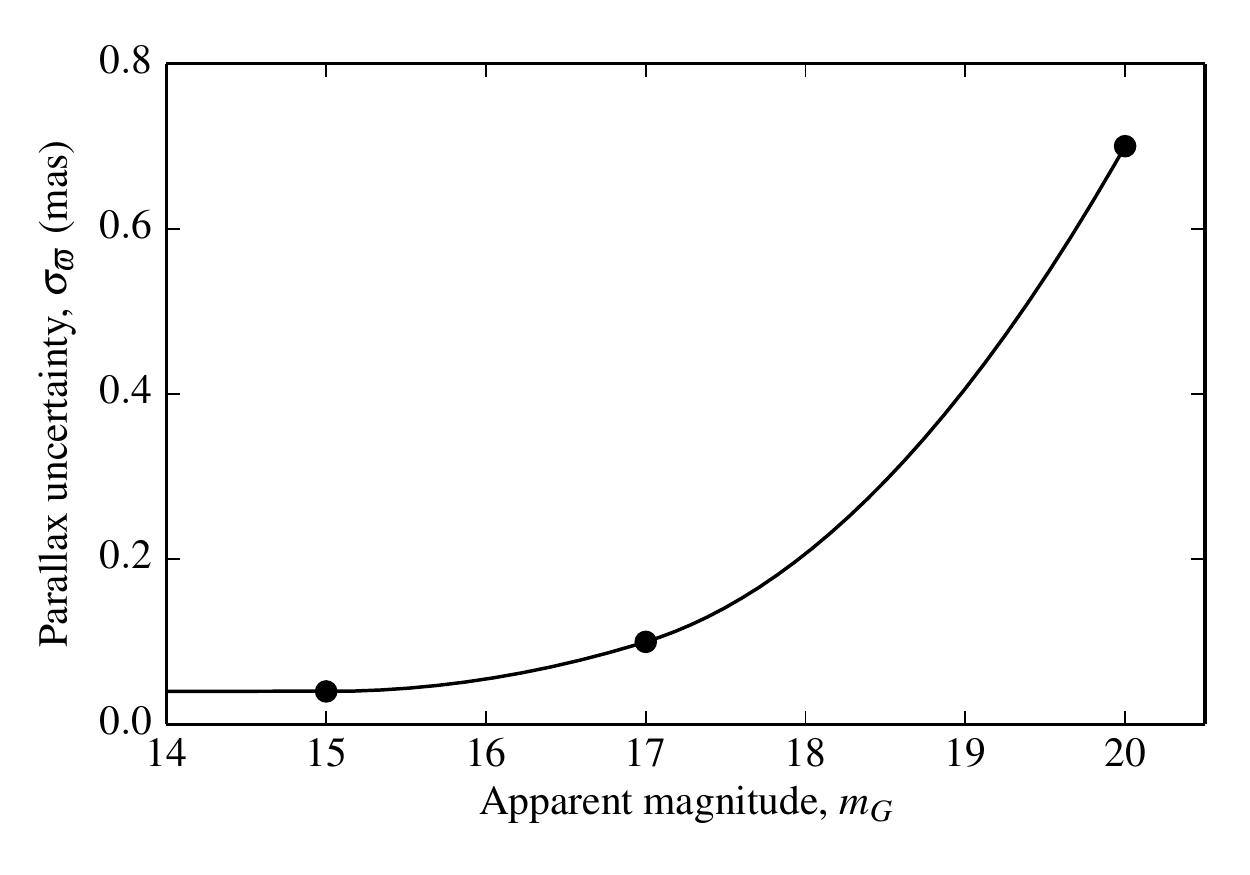}
    \caption{Parallax uncertainties as a function of \emph{Gaia} $G$-band apparent magnitude $m_G$. The dots correspond to magnitudes $m_G=\{15,17,20\}$. For $m_G\leq 15$ the parallax uncertainty is 0.04 mas and constant. At higher $m_G$, the dots are interpolated using a second order polynomial spline.}
    \label{fig:parallax_error}
\end{figure}

We use parallax information with the precision of \emph{Gaia} DR2, which has a limiting magnitude of $m_G \simeq 21$. As listed in \cite{2018arXiv180409366L}, the parallax uncertainties of \emph{Gaia} DR2 are about $0.04$ mas for sources with $m_G<15$, about $0.1$ mas for sources with $m_G=17$, and about $0.7$ mas at $m_G=20$. In order to account for this magnitude dependence, we interpolate these points as shown in Fig. \ref{fig:parallax_error}. The errors in the \emph{Gaia} photometric $G$-band are interpolated in the same manner as for the parallax, with errors of 0.3 mmag for $m_G = 13$, 2 mmag for $m_G = 17$, and 10 mmag for $m_G = 20$.

\section{Statistical model}\label{sec:method}

\begin{figure}
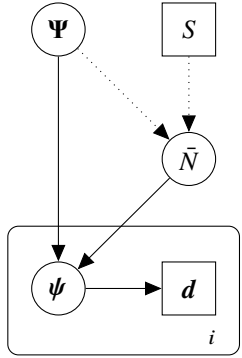
\label{fig:DAG}
  \centering
  \tikz{
    \node[latent] (popp) {$\boldsymbol{\Psi}$} ; %
    \node[latent, right=of popp, rectangle] (selection) {$S$} ; %
    \node[latent, below=of selection] (norm) {$\bar{N}$} ; %
    \node[latent, below=of norm, rectangle] (data) {$\boldsymbol{d}$} ; %
    \node[latent, left=of data] (objp) {$\boldsymbol{\psi}$} ; %
    \plate[inner sep=0.3cm, xshift=-0.0cm, yshift=0.16cm] {index_i} {(data) (objp)} {$i$}; %
    \edge[dotted] {popp} {norm} ; %
    \edge[dotted] {selection} {norm} ; %
    \edge {norm} {objp} ; %
    \edge {popp} {objp} ; %
    \edge {objp}{data}  ; %
  }
  \caption{A directed acyclical graph (DAG) of our statistical model. Quantities inscribed in circles (squares) represent variables (constants); solid (dotted) arrows represent probabilistic (deterministic) dependence; a rectangle with rounded corners represents a set of objects, in this case the set of WD included in our sample, carrying an index $i$; $\popp$ is the set of population parameters, $S$ is the selection function of our sample construction, $\bar{N}$ is the normalization to the WD distribution function, $\objp$ is the set of object parameters, $\boldsymbol{d}$ is the object data, and $i$ is the object index.}
\end{figure}

A directed acyclical graph (DAG) of our statistical model is shown in Fig.~\ref{fig:DAG}. In the DAG, the constants and parameters of the model are inscribed in squares and circles; the arrows indicate the generative relationship between these quantities.

By Bayes' Theorem, the full posterior on both population parameters and object parameters is written
\begin{equation}\label{eq:fullposterior}
	\pr(\popp,\objp | \data ) = \pr(\popp)
    \prod_i \frac{S(\data_i) \pr(\data_i|\objp_i) \pr(\objp_i | \popp)}{\bar{N}(S,\popp)},
\end{equation}
where $\pr(\popp)$ is a prior on the population parameters; $S(\data_i)$ is the probability of being selected given the data of that object; $\pr(\data_i|\objp_i)$ is the likelihood of the data of an object given its object parameters; $\pr(\objp_i | \popp)$ is the probability of object parameters given the population parameters; finally, $\bar{N}(S,\popp)$ is the normalization of $\pr(\objp_i | \popp)$, and depends on the selection function and the population parameters. When writing the data or object parameters without an index $i$, we refer to the complete set of objects, $\objp \equiv \{ \objp_1,\objp_2,...,\objp_N \}$.

\subsection{Object parameters}\label{sec:objectparams}

\begin{figure}
	\includegraphics[width=\columnwidth]{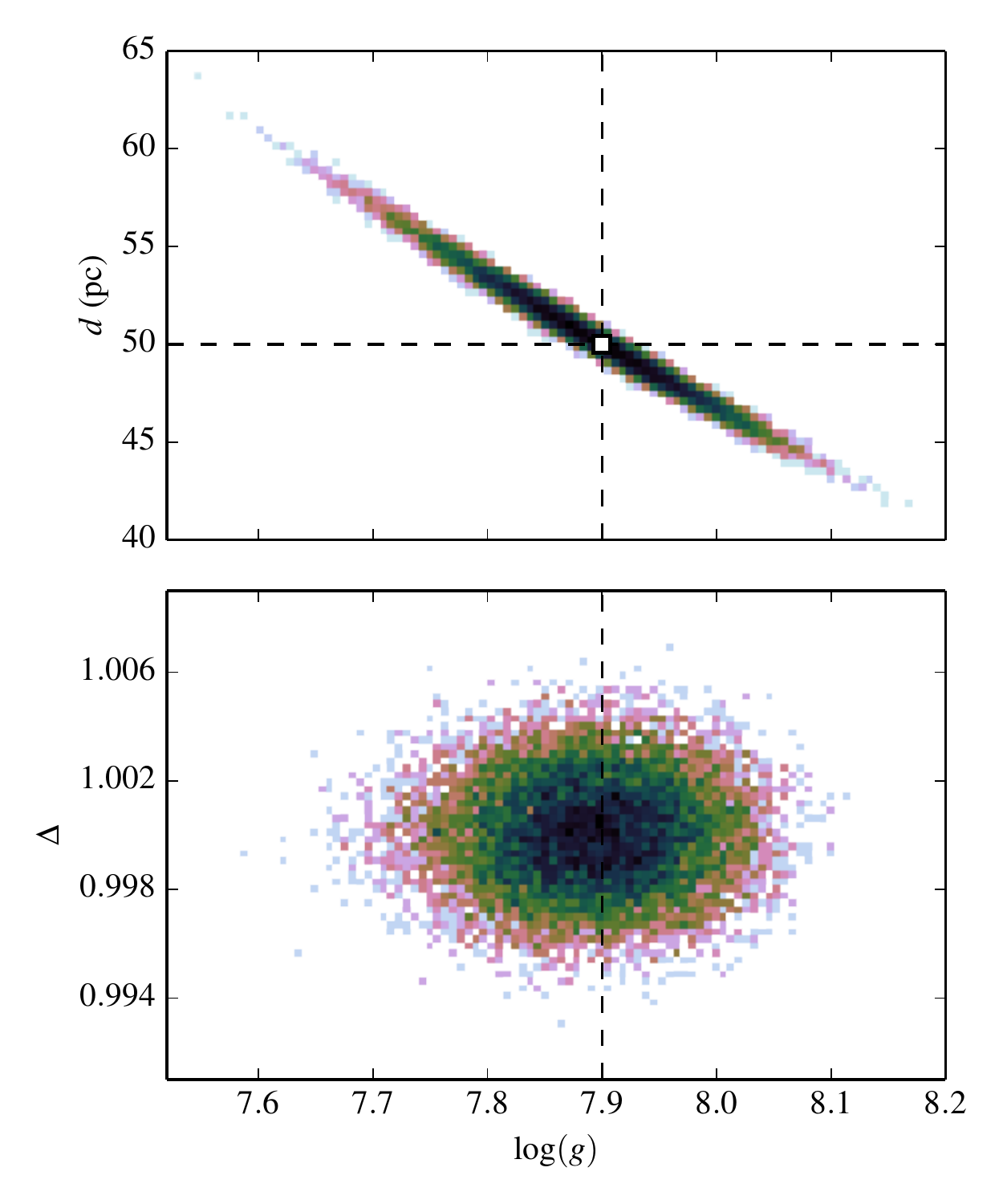}
    \caption{A posterior density of an object with true parameters $\Teff=14000$ K, $\logg=7.9$, $t=\text{DA}$, and $d=50$ pc, with photometric errors of $\sigma_c=0.01$ in all \emph{ugriz} colour bands. The population parameters are set to $\alpha=3.5$, $\bar{g}=7.9$, $\sigma_g=0.1$, $\gamma_g=1.2$, and $f_\text{DB}=0$. Because there is no parallax information, the constraint on the surface gravity is largely due to the prior set by the population model. The two panels show the same posterior density, sharing the surface gravity $\logg$ on the horizontal axis. The top panel shows the distance $d$ on the vertical axis, while the bottom panel shows an alternative parametrization of the distance $\Delta$, described in Sec.~\ref{sec:objectparams}. The true object parameter values are marked with dashed lines and a white square. The correct value for $\Delta$ varies with $\Teff$, which is marginalized over in this figure; hence, there is no true value for $\Delta$, although it should lie close to unity. While the highly degenerate posterior distribution of the top panel can lead to sampling difficulties, the alternative parametrization of the bottom panel avoids such issues.}
    \label{fig:banana}
\end{figure}

Each WD has a set of object parameters encapsulated in $\objp_i$, as listed in Table \ref{tab:parameters}. Because the angular position errors can be neglected, the spatial position only varies in terms of the object's distance. Conceptually, the most straightforward parametrization would be to have the distance $d$ as an object parameter. However, this creates some sampling difficulties arising from the fact that $\logg$ and $d$ are highly degenerate, especially when there is no parallax information available. In this work, we sample the object parameter posteriors using a Metropolis-Hastings Markov Chain Monte-Carlo algorithm \citep{1953JChPh..21.1087M,brooks2011handbook}, which is more efficient when the posterior closer to a multivariate Gaussian in shape. This can be accomplished by a coordinate transformation, as is illustrated in Fig.~\ref{fig:banana}. In the upper panel, the distance is parametrized in terms of $\Delta$, which is the relative change to the ideal distance given $\Teff$ and $\logg$. Let $\tilde{d}(\Teff,\logg)$ be the distance that maximizes the colour factor of the likelihood function, which is
\begin{equation}
	\tilde{d} = 
    h^{-1}\left\{ \frac{\sum_b \sigma_b^{-2} [\hat{m}_c-M_c(\Teff,\logg)]}{\sum_b \sigma_b^{-2}} \right\}.
\end{equation}
The function $h^{-1}$ is the inverse of $h(d)=5[\log_{10}(d/\text{pc})-1]$, the difference between apparent and absolute magnitude. The distance parametrized by the object parameters is $d=\Delta\tilde{d}(\Teff,\logg)$. It is crucial to account for the Jacobian factor that arises with this parametrization, in which the differentials are replaced according to
\begin{equation}
	\de \Teff\, \de \logg\, \de d \rightarrow \de \Teff\, \de \logg\, \de \Delta\, J(\Teff,\logg),
\end{equation}
where the Jacobian is
\begin{equation}
	J(\Teff,\logg) = \tilde{d}(\Teff,\logg),
\end{equation}
which has no dependence on $\Delta$.

\subsection{Object posterior}\label{sec:objectposterior}

The posterior on a specific object also includes the term $\pr(\objp_i | \popp)$, normalized by the quantity $\bar{N}(S,\popp)$. It is given by the population parametrized by
\begin{equation}
\begin{split}
	& \pr(\objp_i | \popp)\, \de \Teff\, \de \logg\, \de d  \\ & \propto
    d^2 n(l,b,\varpi)\, \pr(\Teff,g,t | \popp)\, \de \Teff\, \de \logg\, \de d \\
    & = \Delta^2\, \tilde{d}^3(\Teff,\logg,t) n(l,b,\varpi)\, \pr(\Teff,g,t | \popp)\, \de \Teff\, \de\logg\, \de \Delta,
\end{split}
\end{equation}
where $n$ is the number density of WDs as a function of spatial position. It is implicit in this expression that the true parallax $\varpi$ is a function of the object parameters $\objp_i$.

The normalization factor is given by an integral (or sum, in the case of a discrete variable) over the possible properties of a hypothetical WD drawn from the population model, multiplied by the probability of being selected, as
\begin{equation}\label{eq:normalization}
	\bar{N}(S,\popp) = \sum_{t} \int \de\Teff\, \de \logg\, \de^3\mathbf{x}\,
    \pr(\Teff,g,t | \popp)\, n(\mathbf{x})\, S(\Teff,\logg,t,\mathbf{x}).
\end{equation}
The selection function, $S(\Teff,\logg,t,\mathbf{x})$, is the probability of being included in the sample, given by an object's intrinsic properties and the sample construction cuts on observables.

\section{Mock data and analysis}\label{sec:mock}

To test the algorithm, mock data are generated from the population model. While the exact values of the population parameters are of lesser importance (the main focus being the statistical method), we chose values with the following motivations.

\begin{itemize}
	\item For the temperature distribution we chose $\alpha=3.5$. To first order, WDs cool at a rate of $\dot{\Teff} = \Teff^{-3.5}$, according to \cite{1952MNRAS.112..583M}, although numerical models differ from this cooling rate especially for cooler WDs.
	\item For the distribution of surface gravity, we chose $\bar{g}=7.9$, $\sigma_g=0.1$, and $\gamma_g=1.2$. The surface gravity distribution mean and dispersion ($\bar{g}$ and $\sigma_g$), motivated by \cite{2006ApJS..167...40E}. The tails of the surface gravity distribution are observed to be heavier than those of a Gaussian distribution, motivating a value $\gamma_g>1$.
	\item For the fraction of DB WDs, we chose $f_\text{DB}=0.1$. The fraction of DB WDs is observed to vary with temperature, roughly in the range of 10--20 per cent \citep{2011ApJ...737...28B}.
\end{itemize}

We compare the case of having no astrometric information, versus having parallax measurements with the precision of \emph{Gaia} DR2.

\subsection{Sample selection}\label{sec:sample_cuts}

We define a sample by making cuts on observable properties, in our case on mock data of a generated catalogue. Depending on the errors of these observables, these cuts correspond more or less well to cuts in the objects' intrinsic properties. We do not make a volume-limited sample by making cuts on parallax -- we wish to compare with the case where astrometry is not available, hence we need to construct the sample without such information. We make cuts in observed apparent magnitude and observed colour. The cuts in colour correspond to upper and lower limits on the temperature of WDs included in our sample. The limit in apparent magnitude sets a maximum distance for a WD, as a function of temperature, surface gravity and type.

There are several reasons for not allowing very high temperatures in our samples (although the exact limit is rather arbitrary). Very hot objects are rare in terms of spatial density, but because they are seen to much greater distances they can still be numerous in a sample that is not volume-limited. How many are seen depends on the properties of the population, but this is degenerate with the assumed spatial distribution and the distribution of Galactic dust. Furthermore, with sufficiently high temperature, the peak of an object's spectrum is of shorter wavelength than the $u$ band, in which case the inference on an object's temperature becomes very weak. When working with actual data, it is also necessary to identify what objects are WDs and what objects are not. With very hot, far away objects this is difficult, especially since the distance will be poorly constrained. These issues can be circumvented with good parallax measurements, enabling the construction of a volume-limited sample.

\begin{figure}
	\includegraphics[width=\columnwidth]{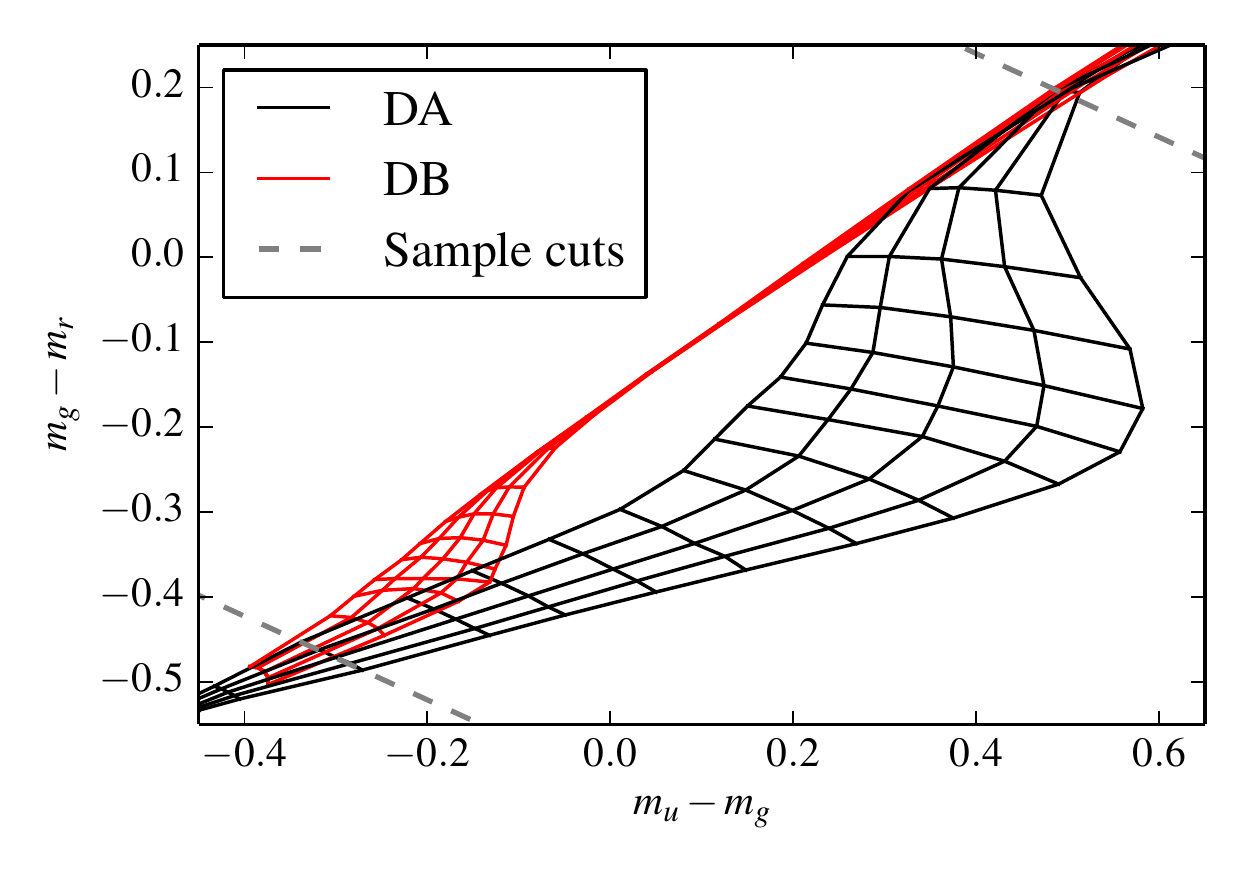}
    \caption{Colours of a DA and DB WD, in contours of constant $\Teff$ or $\logg$. The surface gravity takes values $\logg = \{7,7.5,8,8.5,9\}$. The dashed lines corresponds to the colour cuts in $\hat{\delta}_{ugr}$, where only objects that fall in the region between these lines are included. Because this is a cut in observed colour, observational errors can scatter objects across the sample boundary.}
    \label{fig:colors_cut}
\end{figure}

We make the following cut in colour, demanding that
\begin{equation}
	-0.6195 < \hat{\delta}_{ugr} < 0.4369,
\end{equation}
where
\begin{equation}
	\hat{\delta}_{ugr} \equiv -0.4925\hat{m}_u-0.5075\hat{m}_g+\hat{m}_r.
\end{equation}
Without measurement errors, this cut corresponds to limiting the temperature of a DA type WD to $\Teff \in (7000,30000)$ K; for a DB WD the upper limit in temperature is less restrictive, as can be seen in Fig.~\ref{fig:colors_cut}.

We also make a cut in brightness, by demanding that the \emph{Gaia} $G$ band apparent magnitude fulfils $\hat{m}_G < 20$. In principle, this criterion could equally well have been formulated in terms of some combination of the $ugriz$ apparent magnitudes.

Given these cuts on observables, the selection function is
\begin{equation}\label{eq:selection}
\begin{split}
	S(\Teff,\logg,t,\mathbf{x}) = 
    	      \int \de \hat{m}_G \Theta \big( 20-\hat{m}_G \big)\mathcal{N}\big( \hat{m}_G | m_G,\sigma_G \big) \times \\
    \int \de \hat{\delta}_{ugr}\,
    \Theta \big( \hat{\delta}_{ugr} -0.6195 \big)
    \Theta \big( 0.4369 - \hat{\delta}_{ugr} \big)
    \mathcal{N}\big( \hat{\delta}_{ugr} | \delta_{ugr},\sigma_{\delta}\big),
\end{split}
\end{equation}
where $m_G$ and $\delta_{ugr}$ are true observables with an implicit dependence on the object parameters. The error on $\hat{\delta}_{ugr}$ is given by
\begin{equation}
	\sigma_\delta^2 = (0.4925 \sigma_u)^2 + (0.5075 \sigma_g)^2 + \sigma_r^2,
\end{equation}
assuming that the different magnitude errors are uncorrelated.

\subsection{Generating a mock sample}

The mock sample of WDs is generated by rejection sampling. An object is drawn randomly from the true population model: the object parameters $\Teff$, $\logg$ and $t$ are distributed as described in Eq.~\eqref{eq:T&g} and Eq.~\eqref{eq:DADB} and can be randomized analytically; the position is distributed according to $n(\mathbf{x})$ and is randomized by rejection sampling, knowing that there is a maximal distance a WD can have in order to be included in our sample (observational errors included). A randomly drawn object is then assigned observable quantities, with errors as described in Sec.~\ref{sec:data}. If the object observables fulfil the selection cuts, the object is included in the sample; if not, it is rejected.

\begin{figure*}
	\includegraphics[width=.9\textwidth]{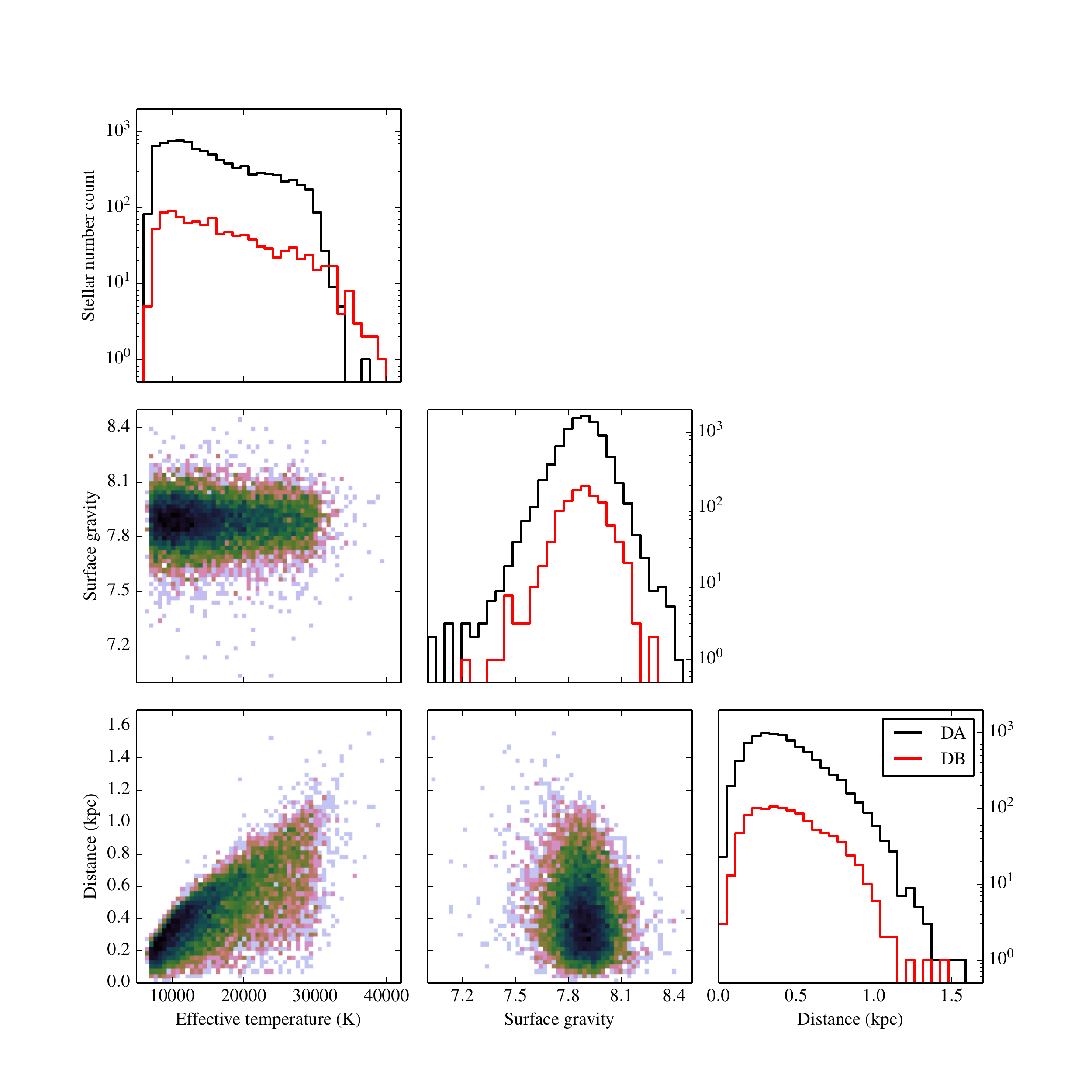}
    \caption{The distributions of true intrinsic properties of our mock data sample, represented as 1D and 2D histograms. The axis are shared between the panels, except for the vertical axis of the 1D histograms. DA and DB WDs are plotted separately in the 1D histograms, but not in the 2D histograms.}
    \label{fig:10000WDs}
\end{figure*}

We construct a sample with 10,000 WDs. The distribution of true object parameter values is shown in Fig. \ref{fig:10000WDs}, where selection effects are manifest. The maximum distance is clearly seen as a function of temperature, where hotter objects are seen further away. Due to the colour cut, the high temperature tail is more pronounced for the DB sub-population. It is also clear that low mass WDs are more likely to be included as they are more luminous, giving rise to some asymmetry in the distribution of surface gravity. The hottest object in this sample has an effective temperature of $\Teff=39,060$~K and is of DB type. The most distant object is located at 1.56~kpc, is of DA type and has an effective temperature $\Teff=37,287$~K surface gravity $\logg=7.72$.

\subsection{Results}

We infer the population parameters of a Bayesian hierarchical model, as described in Sec.~\ref{sec:model}, for our mock data sample, using a Monte-Carlo Markov Chain (MCMC) to trace the posterior distribution. We use a special purpose sampling algorithm called Metropolis-within-Gibbs \citep{BayesianDataAnalysis}, in which the population parameters ($\popp$) and the object parameters of each individual WD ($\objp_i$) are varied separately, producing a computationally efficient marginalization of the object parameters. The WD type ($t$), which is a discrete variable, alternates at random for each attempted step of the object parameter Metropolis step. For all other variables, which are continuous, the step is drawn at random from a multivariate normal distribution. The covariance matrices of these multivariate Gaussians, one for the population parameters and one for each of the 10,000 objects, are tuned in a thorough burn-in phase. For each iteration of the algorithm, the algorithm attempts 40 steps in population parameter space, and 40 steps in the parameter space of each separate object.

The population parameter prior, $\pr(\popp)$, is taken to be uniform and wide in all parameters around the true parameter values, with the exception of the lower bound to $\gamma_g$ which is set to 1.

\begin{figure*}
	\includegraphics[width=1.\textwidth]{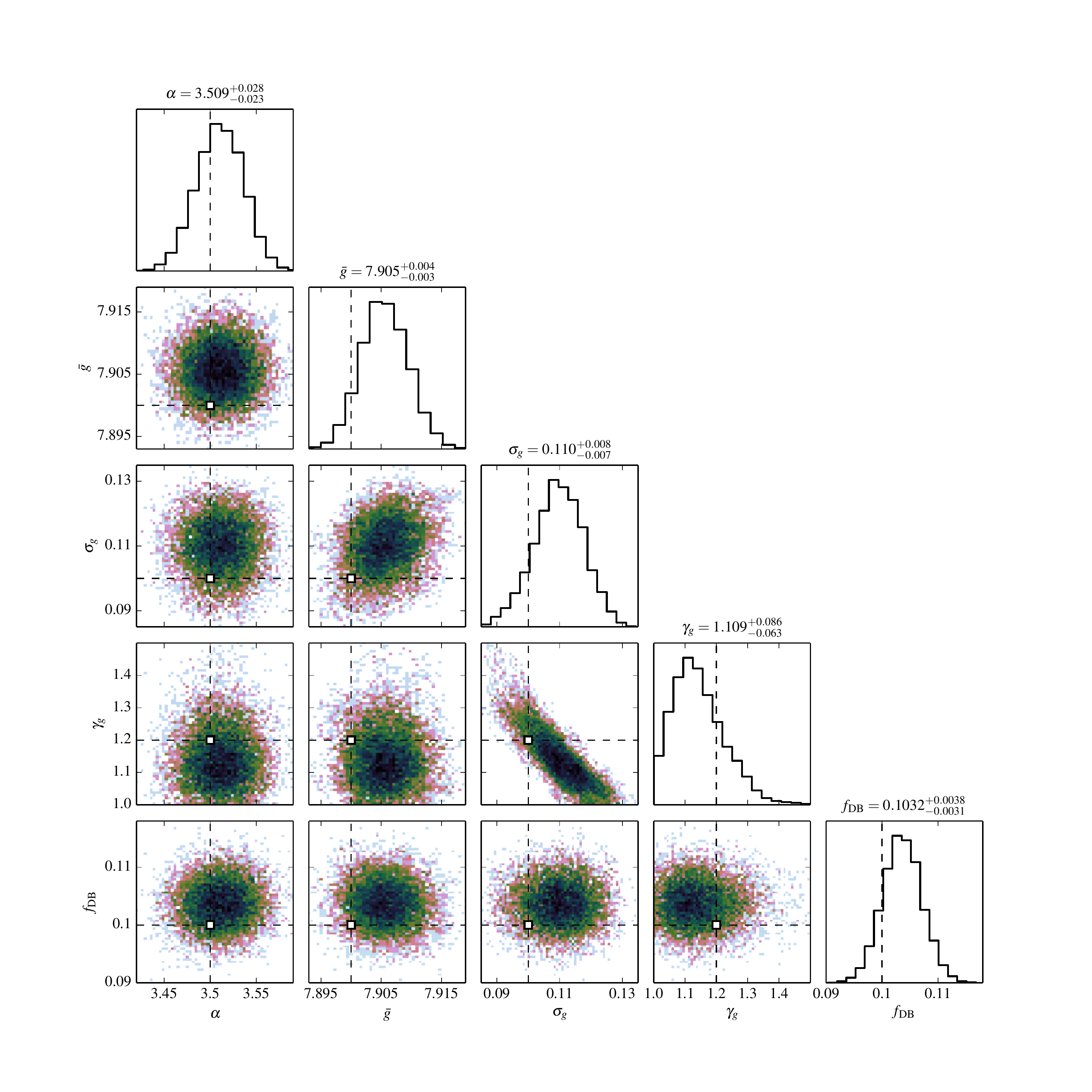}
    \caption{ Posterior density of the population parameters, for a mock data set with no astrometric information. The highest posterior density (HPD) credible interval, presented in terms of the HPD value plus/minus bounds that cover 68 per cent of the posterior density, are indicated. The true population parameter values are marked with dotted black lines, and a white square in the 2D histograms.}
    \label{fig:chain}
\end{figure*}

\begin{figure*}
	\includegraphics[width=1.\textwidth]{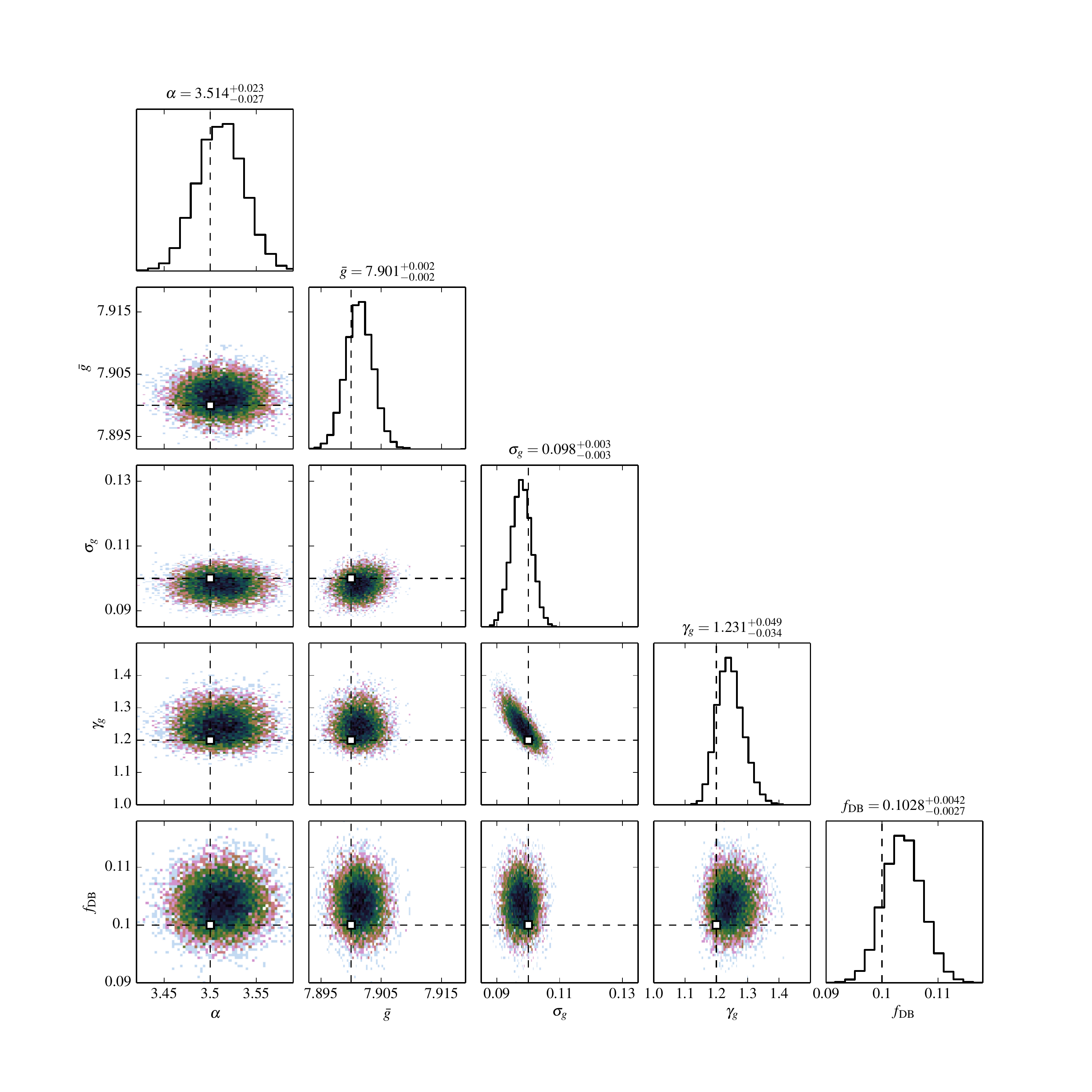}
    \caption{ Posterior density of the population parameters, for a mock data set with parallax information. The highest posterior density (HPD) credible interval, presented in terms of the HPD value plus/minus bounds that cover 68 per cent of the posterior density, are indicated. The true population parameter values are marked with dotted black lines, and a white square in the 2D histograms. The axis range of all panels are the same as in Fig.~\ref{fig:chain}.}
    \label{fig:chain_parallax}
\end{figure*}

The inferred posterior distributions are shown in Fig. \ref{fig:chain} and \ref{fig:chain_parallax}, where the former has no parallax information. In each of these chains, the Metropolis-within-Gibbs MCMC has run for 10,000 iterations.

In both cases, with and without parallax information, the correct population parameter values are recovered by the posterior distribution. The inference on quantities $\alpha$ (parametrizing the distribution of effective temperature) and $f_\text{DB}$ (the fraction of DB WDs) is not significantly affected by also adding parallax information. For the remaining parameters, $\bar{g}$, $\sigma_g$, and $\gamma_g$ (parametrizing the distribution of surface gravity), the inference is strongly affected when including parallax information. The width of the posterior distribution is roughly halved for all three quantities. There is a clear degeneracy between $\sigma_g$ and $\gamma_g$, following the relation where the total variance on the surface gravity, $\text{Var}(\logg) = \sigma_g^2\gamma_g^2$, is preserved.

In Table~\ref{tab:posteriors}, we present highest posterior density (HPD) intervals for the population parameters of three mock data samples, the first one being the sample discussed above. There is no indication of any bias in the inference of the population parameters.

\begin{table*}
	\centering
	\caption{The 68 per cent HPD credible intervals for three different mock samples. The posteriors distribution of sample \textbf{A} is also shown in Fig.~\ref{fig:chain} and \ref{fig:chain_parallax}.}
	\label{tab:posteriors}
    \begin{tabular}{l l l l l l}
		\hline
		Sample  & $\alpha$ & $\bar{g}$ & $\sigma_g$ & $\gamma_g$ & $f_\text{DB}$  \\
		\hline
		\textbf{A}, without parallax &
			$3.509^{+0.028}_{-0.023}$ & 
			$7.905^{+0.004}_{-0.003}$ & 
			$0.110^{+0.008}_{-0.007}$ & 
			$1.109^{+0.086}_{-0.063}$ & 
			$0.1032^{+0.0038}_{-0.0031}$ \vspace{2pt} \\
		\textbf{A}, with parallax &
			$3.514^{+0.023}_{-0.027}$ &
			$7.901^{+0.002}_{-0.002}$ & 
			$0.098^{+0.003}_{-0.003}$ &
			$1.231^{+0.049}_{-0.034}$ &
			$0.1028^{+0.0042}_{-0.0027}$ \\
		\hline
		\textbf{B}, without parallax &
			$3.500^{+0.023}_{-0.027}$ & 
			$7.900^{+0.004}_{-0.003}$ & 
			$0.092^{+0.009}_{-0.007}$ & 
			$1.261^{+0.104}_{-0.099}$ & 
			$0.1001^{+0.0041}_{-0.0030}$ \vspace{2pt} \\
		\textbf{B}, with parallax &
			$3.501^{+0.023}_{-0.028}$ & 
			$7.903^{+0.002}_{-0.002}$ & 
			$0.101^{+0.003}_{-0.004}$ & 
			$1.206^{+0.046}_{-0.034}$ & 
			$0.0992^{+0.0044}_{-0.0025}$ \\
		\hline
			\textbf{C}, without parallax &
			$3.472^{+0.024}_{-0.026}$ & 
			$7.898^{+0.003}_{-0.004}$ & 
			$0.091^{+0.006}_{-0.010}$ & 
			$1.294^{+0.075}_{-0.135}$ & 
			$0.1019^{+0.0036}_{-0.0034}$ \vspace{2pt} \\
		\textbf{C}, with parallax &
			$3.470^{+0.025}_{-0.026}$ & 
			$7.898^{+0.002}_{-0.002}$ & 
			$0.098^{+0.002}_{-0.004}$ & 
			$1.186^{+0.036}_{-0.035}$ & 
			$0.1019^{+0.0036}_{-0.0034}$ \\
		\hline
	\end{tabular}
\end{table*}

\section{White dwarf subpopulations}\label{sec:subpopulations}

The population of WDs in the Milky Way is richer and more complicated than the model described in Sec.~\ref{sec:model}. While we do consider WD sub-populations in the sense of accounting for the difference between DA and DB type WDs, there are other meaningful ways to construct the population model. While we assume that the distribution of temperatures and surface gravities is the same between DA and DB WDs, derived from the same population parameters, it could be meaningful to have separate sets of population parameters for the two types.

In the same vein, one would expect the disk and halo WD population to have different properties. Furthermore, there is expected to be a sub-population of binary WD systems. We discuss how to model these two cases below.

\subsection{Disk and halo populations}

It would be interesting to see qualitative differences between disk and halo WDs. For example, the kinematically warmer halo population has older stars. The population of very old WDs is scientifically interesting, as it holds information about the star formation and dynamical history of the Milky Way.

In this population model, each sub-population would have its own set of population parameters: $\popp = \{ \popp_\text{disk},\popp_\text{halo} \}$. They would each have their respective spatial number density distributions: $n_\text{disk}(\mathbf{x})$ and $n_\text{halo}(\mathbf{x})$. It would be necessary to have a population parameter that describes the relative number density fraction of the two sub-populations at some reference point (for example the Sun's position), such that they can be normalized. The posterior, analogous to Eq.~\eqref{eq:fullposterior} would read
\begin{equation}\label{eq:posterior_disk_halo}
\begin{split}
	& \pr(\popp,\objp | \data ) = \\
	& \pr(\popp)
	\prod_i 
	\frac{S(\data_i) \pr(\data_i | \objp_i)
	\Big[ \pr(\objp_i | \popp_\text{disk})+\pr(\objp_i | \popp_\text{halo}) \Big] }
	{\bar{N}(S,\popp_\text{disk},\popp_\text{halo})}.
\end{split}
\end{equation}
The total number count is simply the sum over the two sub-populations, like
\begin{equation}
	\bar{N}(S,\popp_\text{disk},\popp_\text{halo})=\bar{N}_\text{disk}(S,\popp_\text{disk})+\bar{N}_\text{halo}(S,\popp_\text{halo}).
\end{equation}

\subsection{Binary population}

Unresolved binary WD systems can be identified using only photometry and astrometry, in a similar way to the method presented in \cite{2018ApJ...857..114W}. For a binary system, the likelihood is the same as in Eq.~\eqref{eq:likelihood}, the difference being that the \emph{ugriz} apparent magnitudes of the two component stars are added together, according to
\begin{equation}
	m_{b,\text{sum}} = - \frac{5}{2}\log_{10}\left( 10^{-\frac{2}{5}m_{b,A}}+10^{-\frac{2}{5}m_{b,B}}  \right),
\end{equation}
where $m_{b,A}$ and $m_{b,B}$ are the $b$-band apparent magnitudes of the two component stars.

The posterior density of a binary system will be written in terms of 7 parameters instead of 4, as we have temperature, surface gravity, and type of the two component stars, and the distance of the binary system.

We construct a population of mock binaries by random pairing of the singles population and the same selection criteria, although we also add a constraint in terms of cooling time of the two component stars. In addition to the effective temperature and surface gravity distributions of the two component stars, as described by Eq.~\eqref{eq:T&g}, the probability of pairing also has a factor
\begin{equation}\label{eq:time_difference}
	\exp\left\{
	-\frac{[t_\text{cool}(\Teff_A,\logg_A,t_A)-t_\text{cool}(\Teff_B,\logg_B,t_B)]^2}{2\times ( 500~\text{Myrs})^2}
	\right\},
\end{equation}
where $t_\text{cool}(\Teff_A,\logg_A,t_A)$ is the cooling time of the $A$ component WD (and equivalently for component $B$), as given by the Bergeron atmospheric model. The chosen time difference of $\sim 500$ million years prohibits the pairing of extremely cool and faint WDs with hotter ones. This is a reasonable assumption, as binary stars are typically born in the same system and with similar properties. Without this constraint, one component star would almost always be extremely faint, making binary identification almost impossible. For reference, the cooling time of a WD with $\Teff=10,000~\K$ and $\logg=7.9$ is roughly 500 Myr.

\begin{figure}
	\includegraphics[width=\columnwidth]{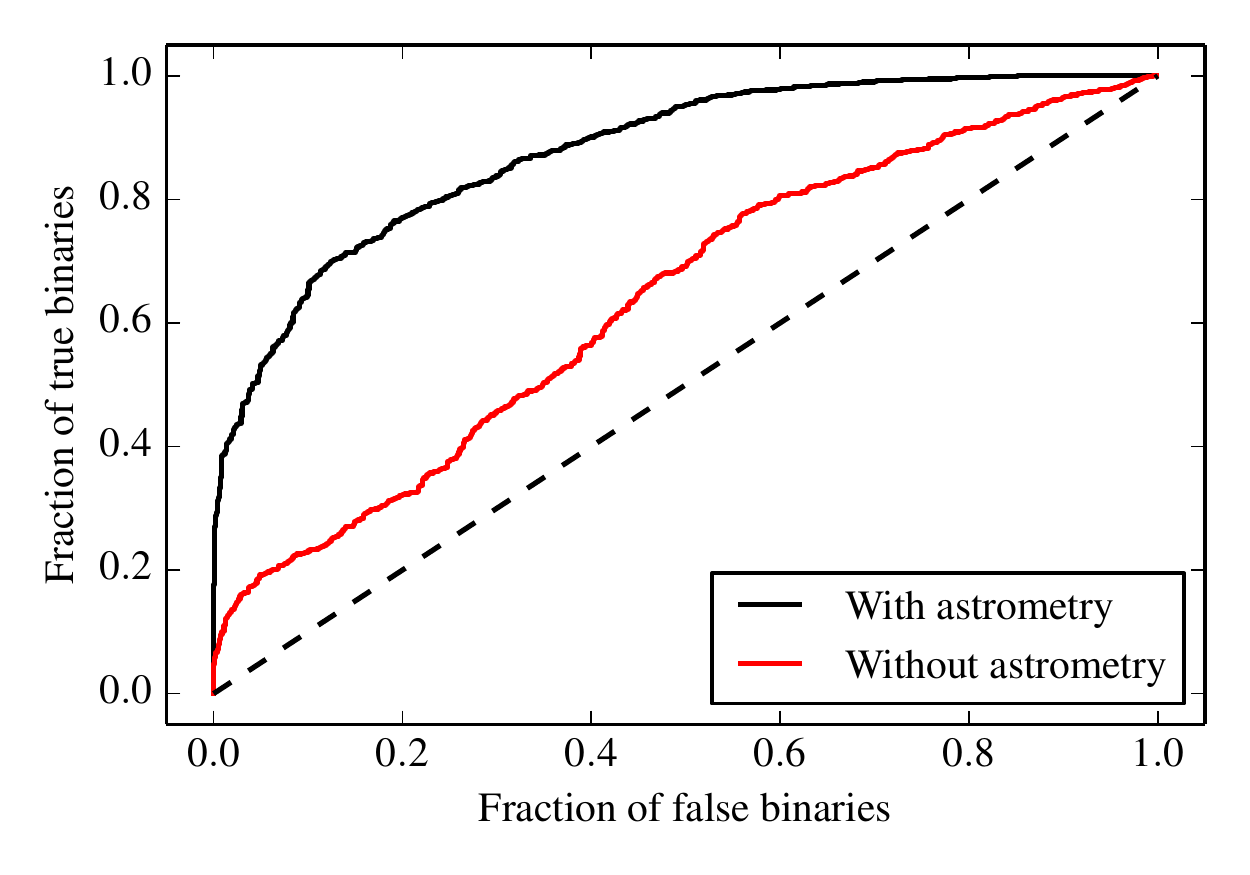}
    \caption{A receiver operating characteristic (ROC) curve of binary WD identification. This shows the rate of falsely identified versus correctly identified binary WD systems, for the cases of having and not having parallax information. An object is labelled binary if the Bayes factor for being a binary system is above some threshold value, where this threshold value is monotonically decreasing along the graphs, from the bottom-left to the top-right corner of the figure. The dashed line shows the linear relationship.}
    \label{fig:ROC_binaries}
\end{figure}

Figure \ref{fig:ROC_binaries} shows a receiver operating characteristic (ROC) curve for identification of binaries, for the cases of having and not having parallax information. The binaries are inferred with knowledge of the underlying population model, in the sense that the population parameters are known. The Bayesian evidence for being a single WD and being a binary WD is computed by sampling the posterior over the object parameters with an MCMC, and then approximating the integral by a first order multivariate Gaussian approximation, given by the covariance matrix and maximum posterior value MCMC chain. The integral is computed for all possible DA and DB combinations separately, with a $\Teff_A>\Teff_B$ multiplicity constraint for binary WDs, circumventing issues of multimodal posterior densities. This is done for 1000 mock data single WDs and 1000 mock data binary WD systems. It is clear from Fig.~\ref{fig:ROC_binaries} that binary identification is significantly improved with parallax information, for which some binaries can be strongly identified even with a very low contamination rate (20 per cent binary identification with 0.1 per cent contamination).

This identification is made on mock data when the underlying population model is known. Working with real data brings many complications, not least coming from the fact that the population model is unknown and inferred. Even so, this test shows that it should be possible to identify a WD binary population. An important aspect that is not accounted for here is that some WD binaries are not drawn from the same distribution of masses as the population of single WDs. A tight binary system goes through phases of mass transfer and shared envelopes; thus there will be binaries with component WD with low mass and surface gravity. Such a binary system would actually be even easier to detect using this method, as they would be brighter due to multiplicity, and brighter still from being low mass and larger in size.

\section{Discussion}\label{sec:discussion}

In this paper, we demonstrate how to infer properties of the local WD population using only astrometric and photometric information, in the framework of a Bayesian hierarchical model.

In our mock sample, we have limited ourselves to a total number of 10,000 WDs and a simple population model, in order to demonstrate the statistical method. The catalogue of WDs in a \emph{Gaia} and SDSS cross-matched sample is expected to be around an order of magnitude larger, enabling us to fit a significantly more complicated model. The model could be extended with more complex distributions of effective temperature, surface gravity, and type, and by including sub-populations as discussed in Sec.~\ref{sec:subpopulations}. With a kinematic model, proper motion information would be very informative, especially in terms of differentiating between disk and halo WDs.

When working with real data, there are complications that are not included here but would be straightforward to implement within this framework. Most WD seen by \emph{Gaia} and SDSS are very close to the Sun and almost unaffected by dust. However, hotter and more luminous WDs are seen to further distance and subject to dust reddening and extinction. With a good dust map, selection effects and photometric reddening for such objects can be accounted for. Also not included in this work are incompleteness effects, which are severe for WDs in \emph{Gaia} DR2. This will improve significantly with future data releases, but will still be crucial to account for.

\emph{Gaia} parallax measurements provide robust identification of WDs, enabling the construction of volume-limited samples, and breaks the degeneracy between distance and size. It is possible to differentiate sub-populations of WDs using this method, such as a population of binary WD systems. Our statistical model fully and correctly accounts for selection effects and observational uncertainties, permitting the construction of a large data sample, without the need to exclude objects with low signal-to-noise or missing parallax information.

\section*{Acknowledgements}

We would like to thank Pierre Bergeron, for providing WD atmospheric models with SDSS and \emph{Gaia} passband photometry. This work was performed in part at the Aspen Center for Physics, which is supported by National Science Foundation grant PHY-1607611. This work was also partially supported by a grant from the Simons Foundation.



\bibliographystyle{mnras}
\bibliography{refs} 





\bsp	
\label{lastpage}
\end{document}